\documentclass[10pt,twocolumn]{IEEEtran}

\usepackage{cite}
\usepackage{graphicx}
\usepackage{epsfig}
\usepackage{psfrag}
\usepackage{subfigure}
\usepackage{url}
\usepackage{stfloats}
\usepackage{amsmath}
\usepackage{color}
\usepackage{amssymb}
\usepackage{epsfig}
\usepackage{epstopdf}

\usepackage{algorithm}
\usepackage{algorithmic}

\usepackage{array}

\begin{document}
%
% paper title
\title{Compressed Sensing of EEG for Wireless Telemonitoring with Low Energy Consumption and Inexpensive Hardware}

\author{Zhilin Zhang, \IEEEmembership{Student Member, IEEE}, Tzyy-Ping Jung, \IEEEmembership{Senior Member, IEEE} \\Scott Makeig, \IEEEmembership{Member, IEEE}, Bhaskar D. Rao, \IEEEmembership{Fellow, IEEE}
\thanks{The work was supported by NSF grants CCF-0830612, CCF-1144258 and DGE-0333451, and was in part supported by Army Research Lab, Army Research Office, Office of Naval Research and DARPA. }
\thanks{Z.Zhang and B.D.Rao are with the Department of Electrical and Computer Engineering, University of California, San Diego, La Jolla, CA 92093-0407, USA. Email:\{z4zhang,brao\}@ucsd.edu}
\thanks{T.-P. Jung and S. Makeig are with the Swartz Center for Computational Neuroscience, University of California, San Diego, La Jolla, CA 92093-0559, USA. Email:\{jung,scott\}@sccn.ucsd.edu}
}

% The paper headers
\markboth{IEEE Transactions on Biomedical Engineering, vol. 60, no. 1, pp. 221-224, 2013}{Zhang \MakeLowercase{\textit{et al.}}: }

% make the title area
\maketitle

\begin{abstract}
Telemonitoring of electroencephalogram (EEG) through wireless body-area networks is an evolving direction in personalized medicine. Among various constraints in designing such a system, three important constraints are energy consumption, data compression, and device cost. Conventional data compression methodologies, although effective in data compression, consumes significant energy and cannot reduce device cost. Compressed sensing (CS), as an emerging data compression methodology, is promising in catering to these constraints. However, EEG is non-sparse in the time domain and also non-sparse in transformed domains (such as the wavelet domain). Therefore, it is extremely difficult for current CS algorithms to recover EEG with the quality that satisfies the requirements of clinical diagnosis and engineering applications. Recently, Block Sparse Bayesian Learning (BSBL) was proposed as a new method to the CS problem. This study introduces the technique to the telemonitoring of EEG. Experimental results  show that its recovery quality is better than state-of-the-art CS algorithms, and sufficient for practical use.  These results suggest that BSBL is very promising for telemonitoring of EEG and other non-sparse physiological signals.
\end{abstract}


\IEEEpeerreviewmaketitle
%\newpage

%============================================================================================
\section{Introduction}
\label{sec:intro}
%============================================================================================

Telemonitoring of electroencephalogram (EEG) via Wireless Body-Area Networks (WBANs) is an evolving direction in personalized medicine and home-based e-Health. In such a system, a WBAN \cite{cao2009enabling} integrates a number of sensors which collect and compress EEG. The compressed EEG is sent to a nearby smart-phone via ultra-low-power short-haul radios (e.g., Bluetooth), and then is transmitted to a remote terminal (e.g., a hospital) via the Internet. In the terminal, the original EEG is recovered by a computer. Equipped with the system, patients need not visit hospitals frequently. Instead, their EEG can be monitored continuously and ubiquitously.

However, there are many constraints that have to be taken into account when designing such a system. The primary one is energy constraint \cite{milenkovic2006wireless}. Due to limitation on battery life, it is necessary to reduce energy consumption as much as possible. Low energy consumption means that a system can use small and light batteries and sensors. Consequently, the light weight of the device  can significantly  improve the comfort level of patients. Besides, low energy consumption means longer battery and sensor lifetime, which reduces operational costs of the system.

Another constraint is that transmitted physiological signals should be largely compressed. This is because the communication capacity of ultra-low-power short-haul radio devices is low and can be stressed especially in some applications using multiple-sensors or high-speed sampling frequency. Besides, a WBAN generally uses a smart-phone as an intermediate transit point. Thus, it is important that data stream does not overwhelm the smart-phone,  disturbing its primary functions such as receiving and making phone calls, playing games, and other mobile-based applications.

The third constraint is hardware costs. Low hardware costs are more likely to make a telemonitoring system economically viable and accepted by individual customers. However, low hardware costs mean that data compression (on sensors) should have low complexity and data recovery (in remote terminals) should not require sensors to pre-process raw signals when collecting them.

It is noted that many conventional data compression methodologies such as wavelet compression cannot satisfy all the above constraints at the same time. It has been shown in \cite{mamaghanian2011compressed} that compared to wavelet compression, Compressed Sensing (CS), when using sparse binary matrices as its sensing matrices, can reduce energy consumption while achieving competitive data compression ratio. Besides, the use of sparse binary matrices means the device cost can be largely reduced \cite{mamaghanian2011compressed,gangopadhyay2011system}. However, current CS algorithms only work well for sparse signals or signals with sparse representation coefficients in some transformed domains (e.g., the wavelet domain). Since EEG is neither sparse in the original time domain nor sparse in transformed domains, current CS algorithms cannot achieve good recovery quality.

To recover EEG signals with high quality that satisfies the needs of practical applications, this study proposes using  Block Sparse Bayesian Learning (BSBL) \cite{zhang2012TSP,zhang2011IEEE} to compress/recover EEG. The BSBL framework was initially proposed for signals with block structure \cite{zhang2012TSP} and has been successfully used for the telemonitoring of fetal ECG \cite{zhang2012TBME}, which has obvious block structure. This study explores the feasibility of using the BSBL technique for EEG, which is an example of a signal with an arbitrary waveform and without distinct block structure. The fidelity of recovered EEG signal is assessed by subsequent signal processing such as independent component analysis.

The rest of the paper is organized as follows. Section \ref{sec:bsbl} briefly introduces the CS model and the BSBL framework. Section \ref{sec:experiments}
presents the results. The last two sections discuss the results and conclude the paper.

%============================================================================================
\section{Compressed Sensing and Block Sparse Bayesian Learning}
\label{sec:bsbl}
%============================================================================================

Compressed Sensing (CS) \cite{Baraniuk2007} is a new data compression paradigm, in which a signal of length $N$, denoted by $\mathbf{x}\in \mathbb{R}^{N \times 1}$, is compressed by a full row-rank random matrix, denoted by $\mathbf{\Phi} \in \mathbb{R}^{M \times N} (M \ll N, \mathrm{Rank}( \mathbf{\Phi})=M)$, i.e.,
\begin{eqnarray}
\mathbf{y}= \mathbf{\Phi} \mathbf{x},
\label{equ:SMV model}
\end{eqnarray}
where $\mathbf{y}$ is the compressed data, and $\mathbf{\Phi}$ is called the \emph{sensing matrix}, which is known to CS algorithms for recovery. CS algorithms use the compressed data $\mathbf{y}$ and the sensing matrix $\mathbf{\Phi}$ to recover the original signal $\mathbf{x}$. Their successes rely on the key assumption that most entries of the signal $\mathbf{x}$ are zero (i.e., $\mathbf{x}$ is sparse). When this assumption does not hold, one can seek a \emph{dictionary matrix}, denoted by $\mathbf{D}\in \mathbb{R}^{M \times M}$, so that $\mathbf{x}$ can be expressed as $\mathbf{x}=\mathbf{Dz}$ and $\mathbf{z}$ is sparse. Then, the model (\ref{equ:SMV model}) can be re-written as
\begin{eqnarray}
\mathbf{y}= \mathbf{\Phi} \mathbf{D}\mathbf{z}.
\label{equ:SMV model2}
\end{eqnarray}
Thus, CS algorithms can first recover $\mathbf{z}$ using $\mathbf{y}$ and $\mathbf{\Phi} \mathbf{D}$, and then recover the original signal $\mathbf{x}$ by $\mathbf{x}=\mathbf{Dz}$.

When CS is used in a telemonitoring system, signals are compressed on sensors according to (\ref{equ:SMV model}). This compression stage consumes on-chip energy of the WBAN. The signals are recovered  by a remote computer according to (\ref{equ:SMV model2}), where the matrix $\mathbf{\Phi}$ is known to a CS algorithm and the matrix $\mathbf{D}$ is determined by a user. This stage does not consume any energy of the WBAN.

CS has several advantages over wavelet compression. In \cite{mamaghanian2011compressed,zhang2012TBME} it is shown that when the sensing matrix $\mathbf{\Phi}$ is a sparse binary matrix, in which most entries are zeros and only few entries are ones, CS algorithms cost less energy but have competitive compression ratio compared to wavelet compression. For example, when compressing a signal of length $N$, CS left-multiplies it by an $M \times N (M \ll N)$ sparse binary matrix. Then amplitudes of the $M$ compressed data are coded. In contrast, in wavelet compression this signal is first left-multiplied by an $N \times N$ wavelet transform matrix with real entries. The wavelet coefficients with large amplitudes are found by a search algorithm, and both their amplitudes and locations are coded. Thus, CS requires fewer code execution on CPU. A second advantage of CS is that its compression operator greatly facilitates hardware design, since the implementation of multiplication with a binary matrix needs only accumulator registers. In wavelet compression, the compression involves multiplications of real numbers, which cannot be implemented by merely accumulator registers.

Despite of these advantages, the use of CS in telemonitoring is only limited to a few types of signals, mainly because most physiological signals like EEG are not sparse in the time domain and not sparse enough in transformed domains. The issue now can be solved by the BSBL framework \cite{zhang2012TSP}.

The BSBL framework was initially proposed for recovering a signal with block structure \cite{zhang2012TSP,zhang2011IEEE}. It assumes the signal $\mathbf{x}$ can be partitioned into a concatenation of non-overlapping blocks, and a few of blocks are non-zero. Thus, it requires users to define the block partition of $\mathbf{x}$. However, it turns out that such user-defined block partition does not need to be consistent with the true block partition of the signal \cite{zhang2012TBME}; in fact, the user-defined block partition can be viewed as a regularization for the estimation of the signal's covariance matrix. Further, in this work we found even if a signal has no distinct block structure, the BSBL framework is still effective.
This makes feasible using BSBL for the CS of EEG and adopting the model (\ref{equ:SMV model2}) for recovery, since EEG has arbitrary waveforms and the representation coefficients $\mathbf{z}$ generally lack block structure (see Fig.\ref{fig:EEG1}).

The BSBL framework has a pruning mechanism, which prunes out blocks in $\mathbf{x}$ (or in $\mathbf{z}$ if using the model (\ref{equ:SMV model2})) when the blocks have very small norms. However, EEG is non-sparse in both the time domain and transformed domains. Therefore, we disabled the pruning mechanism in our experiments.

Currently, there are three algorithms in the BSBL framework. In our experiments we chose a bound-optimization based algorithm, denoted by BSBL-BO. Details on the algorithm and the BSBL framework can be found in \cite{zhang2012TSP}.

%============================================================================================
\section{Experiments of Compressed Sensing of EEG}
\label{sec:experiments}
%============================================================================================

The following experiments\footnote{Experiment codes can be downloaded at: \url{https://sites.google.com/site/researchbyzhang/bsbl}.} compared BSBL-BO with some representative CS algorithms in terms of recovery quality. Because all the CS algorithms adopted the same sensing matrix, they had equal energy consumption. Thus, the comparison of energy consumption is excluded.

Two performance indexes were used to measure recovery quality. One was the Normalized Mean Square Error (NMSE), defined as $\|\widehat{\mathbf{x}}-\mathbf{x}\|_2^2/\|\mathbf{x}\|_2^2$, where $\widehat{\mathbf{x}}$ was the estimate of the true signal $\mathbf{x}$. The second was the Structural SIMilarity index (SSIM) \cite{wang2009mean} for 1-dimensional signals (the length of the sliding window was $100$). SSIM measures the similarity between the recovered signal and the original signal, which is a
better performance index than the NMSE for structured signals. Higher SSIM means better recovery quality. When the recovered signal is the same as the original signal, SSIM = $1$.

The following experiments used the model (\ref{equ:SMV model2}) to recover EEG. In the first experiment $\mathbf{D}$ was an inverse Discrete Cosine Transform (DCT) matrix, and thus $\mathbf{z}$ ($\mathbf{z} = \mathbf{D}^{-1}\mathbf{x}$) are DCT coefficients. In the second experiment $\mathbf{D}$ was an inverse Daubechies-20 Wavelet Transform (WT) matrix, which was suggested in \cite{gangopadhyay2011system} for compressing EEG. In both experiments the sensing matrices $\mathbf{\Phi}$ were sparse binary matrices, in which every column contained $15$ entries equal to $1$ with random locations while other entries were zeros. For BSBL-BO, we defined a block partition, where the starting location of each block was incremented by $24$ (i.e., $1, 25, 49,\cdots$). The maximum number of iterations for BSBL-BO was set to $7$.

\subsection{Experiment 1: Compressed Sensing with DCT}

\begin{figure}[tbp]
\begin{minipage}[b]{.48\linewidth}
  \centering
  \centerline{\epsfig{figure=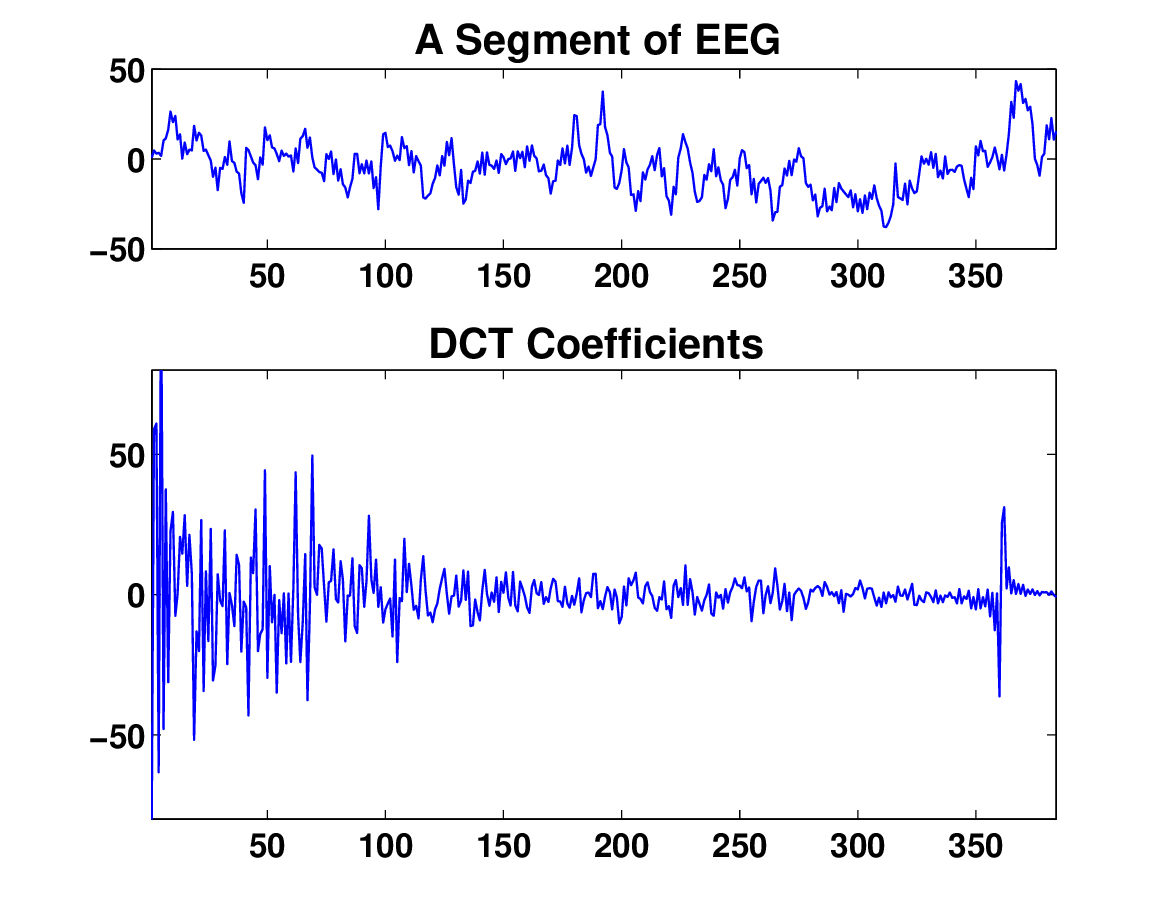,width=4.5cm,height=4.3cm}}
  \centerline{\footnotesize{(a)}}
\end{minipage}
\hfill
\begin{minipage}[b]{0.48\linewidth}
  \centering
  \centerline{\epsfig{figure=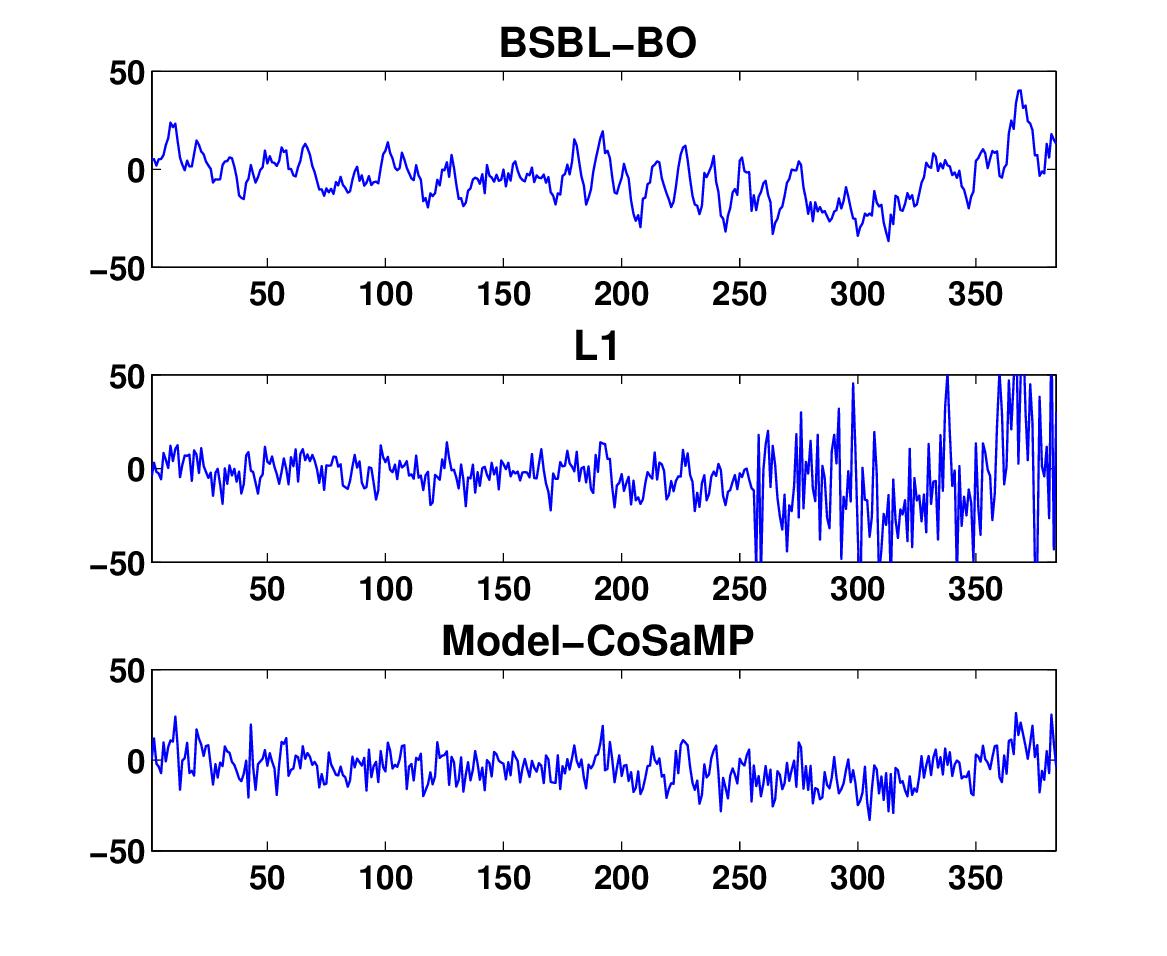,width=4.5cm,height=4.3cm}}
  \centerline{\footnotesize{(b)}}
\end{minipage}
\caption{(a) An EEG epoch, and its DCT coefficients. (b) The recovery results by BSBL-BO, $\ell_1$, and Model-CoSaMP when using the model (\ref{equ:SMV model2}).}
\label{fig:EEG1}
\end{figure}

This example  used a common dataset (`eeglab\_data.set') in the EEGLab \cite{EEGLAB} to mimic the telemonitoring scenario by first compressing it and then recovering it. This dataset contains EEG signals of $32$ channels with
sequence length of $30720$ data points, and each channel signal contains $80$
epochs each containing $384$ points. Artifacts caused by muscle movement are also contained in the signals.

To compress the signals epoch by epoch, we used a $192 \times 384$ sparse binary matrix as the sensing matrix $\mathbf{\Phi}$, and a $384 \times 384$ inverse DCT matrix as the dictionary matrix $\mathbf{D}$.

Two representative CS algorithms were compared in this experiment. One was the Model-CoSaMP \cite{MCS}, which has high performance for signals with known block structure. Here it used the same block partition as BSBL-BO. The second was an $\ell_1$ algorithm used in \cite{gangopadhyay2011system} to recover EEG. The parameters of the two algorithms were tuned for optimal results.

Figure \ref{fig:EEG1}(a) shows an EEG epoch and its DCT coefficients. Clearly, the DCT coefficients were not sparse and had no block structure. Figure \ref{fig:EEG1}(b) shows the recovery results of the three algorithms. Only BSBL-BO recovered the epoch with good quality; characteristic EEG peaks/troughs and oscillatory activities were accurately presented in the recovered signal. Table \ref{Exp1} shows the averaged NMSE and SSIM of the three algorithms on the whole dataset. It also lists the results when BSBL-BO directly recovered the signals without using the dictionary matrix (i.e., using the model (\ref{equ:SMV model})). The DCT-based BSBL-BO evidently had the best performance, and it took 0.105 second per epoch on average on a computer with 2.8G CPU and 6G RAM. BSBL-BO without using the dictionary matrix took 0.271 second per epoch on average.

\begin{table}[tp]
\caption{Averaged performance in Experiment 1. }
\label{Exp1}
\centering
\begin{tabular}{c|m{2.1cm}|m{2.1cm}}
\hline\hline
              &   NMSE (mean $\pm$ std)           &  SSIM (mean $\pm$ std)     \\
\hline
DCT-based BSBL-BO     &   \textbf{0.078 $\pm$ 0.046}         &    \textbf{0.85  $\pm$  0.08}  \\
\hline
BSBL-BO without DCT             &  0.116 $\pm$ 0.066                    &   0.81 $\pm$ 0.09 \\
\hline
DCT-based $\ell_1$      &   0.493 $\pm$ 0.121                   &     0.48 $\pm$ 0.11           \\
\hline
DCT-based Block-CoSaMP  &   0.434 $\pm$ 0.070                  &  0.45 $\pm$ 0.10               \\
\hline
\hline
\end{tabular}
\end{table}

In EEG analysis, a regular methodology is performing Independent Component Analysis (ICA) on scalp EEG data and then analyzing single-trial ERPs for each Independent Component (IC) \cite{jung2001imaging}. Therefore, it is important to examine whether the obtained ICs from the recovered EEG dataset by BSBL-BO are the same as those from the original dataset\footnote{We only need to pay attention to the ICs with large energy, since in regular ICA analysis of EEG, ICs with large energy are reliable and meaningful.}.

This study performed ICA decomposition on the original EEG dataset and the recovered EEG dataset by BSBL-BO, respectively, using the Extended-Infomax algorithm with the same initialization, which is a build-in program in the EEGLab \cite{EEGLAB}. Then, we calculated the back-projected scalp map, the ERP image \cite{jung2001imaging}, and the averaged ERP of each IC from the original dataset and the reconstructed dataset.

\begin{figure}[tbp]
\begin{minipage}[b]{.48\linewidth}
  \centering
  \centerline{\epsfig{figure=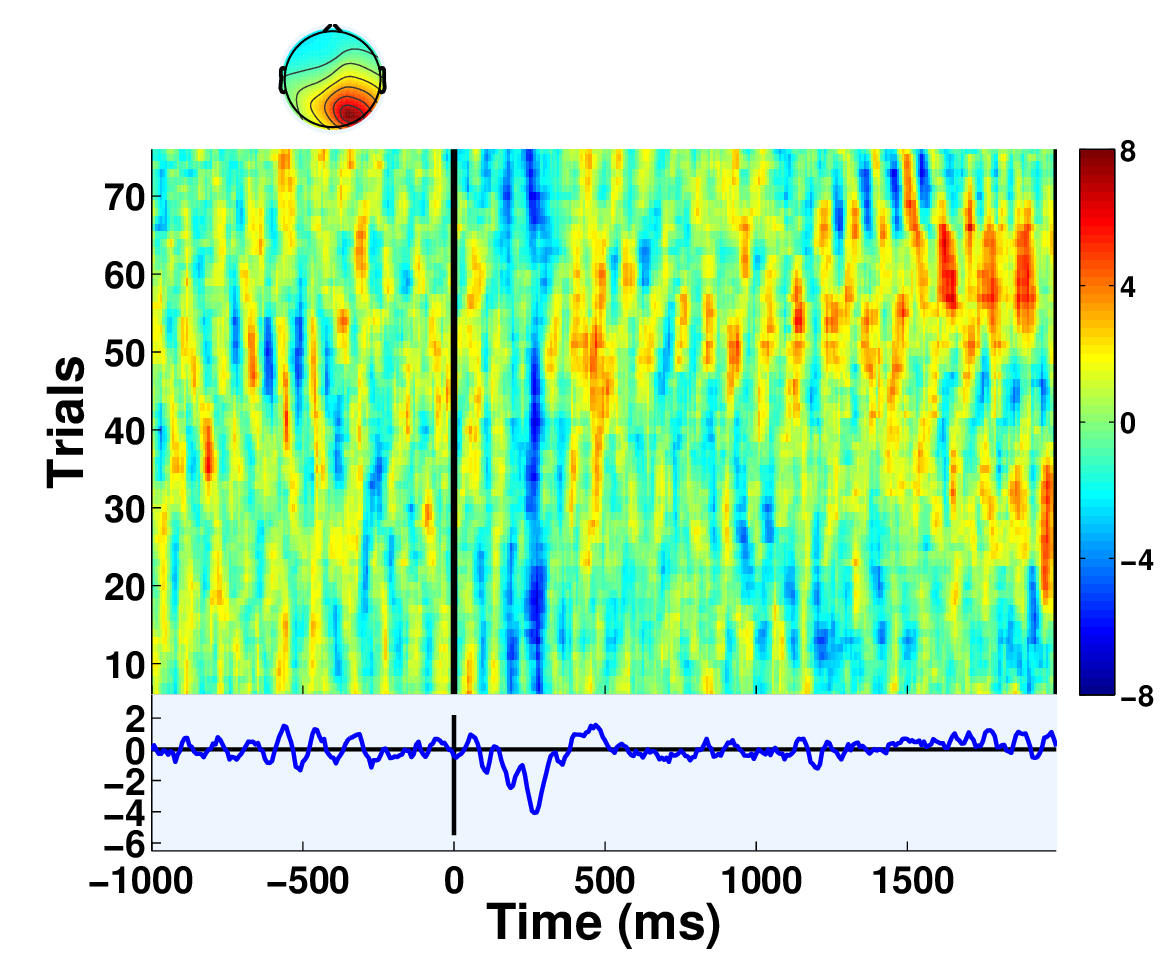,width=4.4cm,height=3.6cm}}
  \centerline{\footnotesize{(a)}}
\end{minipage}
\hfill
\begin{minipage}[b]{0.48\linewidth}
  \centering
  \centerline{\epsfig{figure=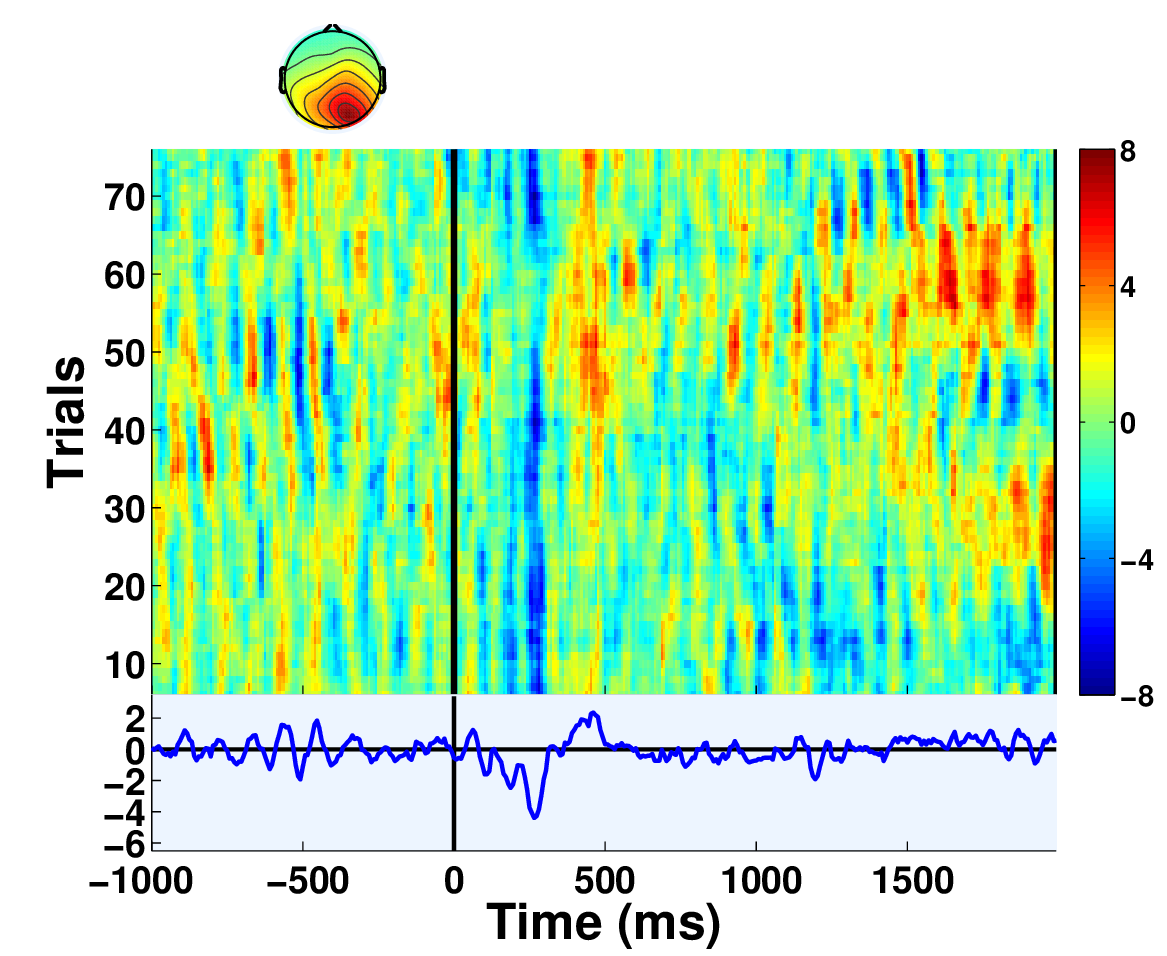,width=4.4cm,height=3.6cm}}
  \centerline{\footnotesize{(b)}}
\end{minipage}
\hfill
\begin{minipage}[b]{.48\linewidth}
  \centering
  \centerline{\epsfig{figure=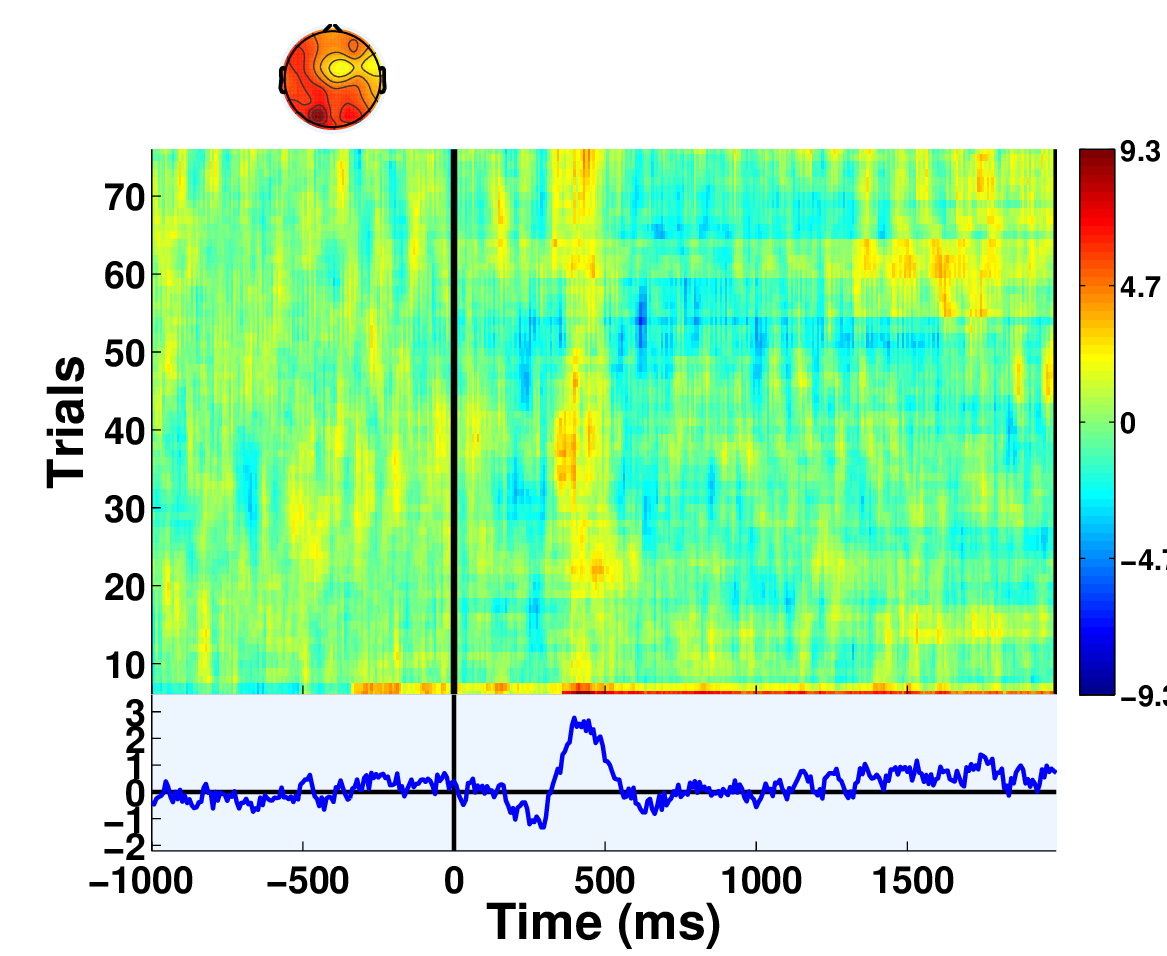,width=4.4cm,height=3.6cm}}
  \centerline{\footnotesize{(c)}}
\end{minipage}
\hfill
\begin{minipage}[b]{0.48\linewidth}
  \centering
  \centerline{\epsfig{figure=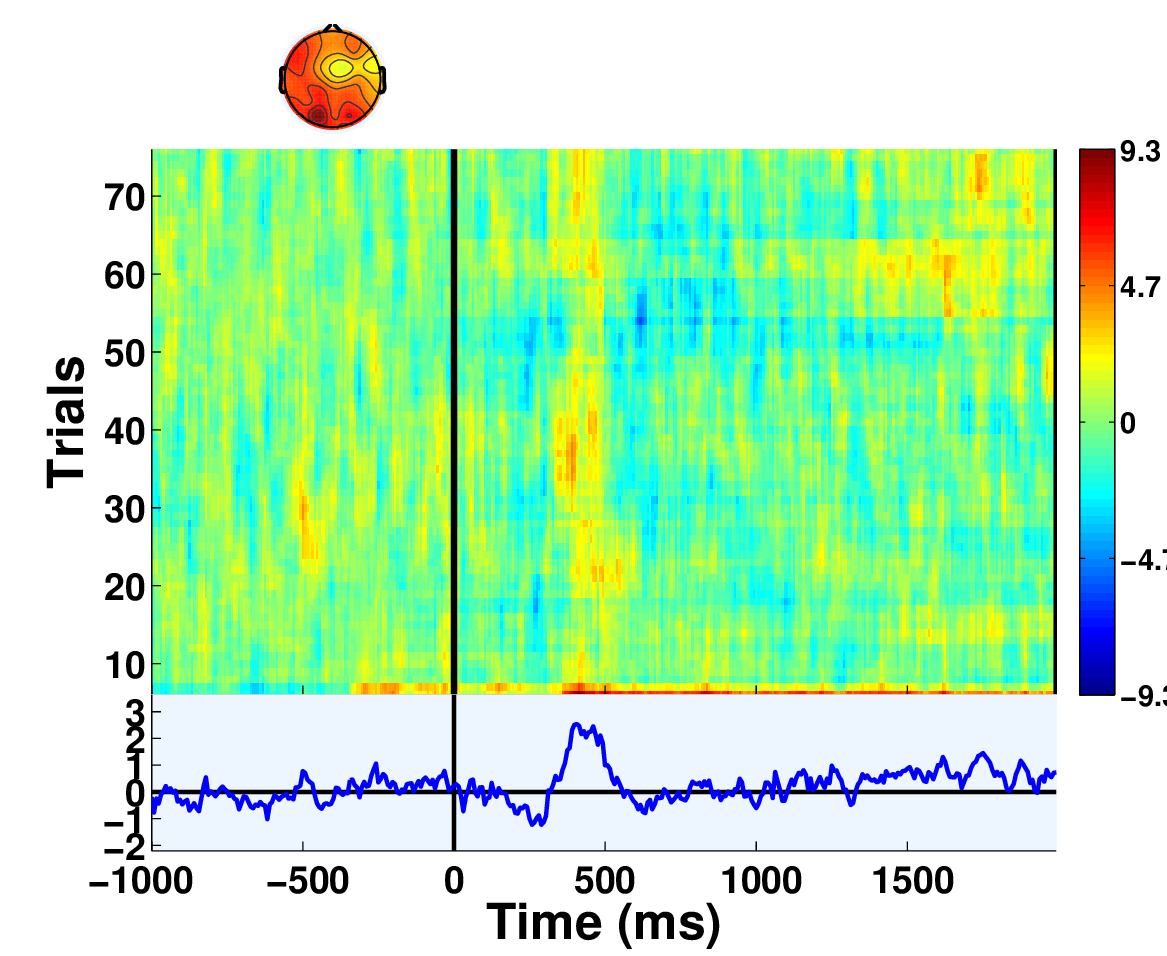,width=4.4cm,height=3.6cm}}
  \centerline{\footnotesize{(d)}}
\end{minipage}
\caption{An IC with focal back-projected scalp distribution derived (a) from the original EEG dataset and (b) from the recovered dataset. Another IC with dispersive scalp distribution derived (c) from the original EEG dataset and (d) from the recovered dataset. Each subfigure shows the back-projected scalp map, the ERP image, and the averaged ERP of an IC.}
\label{fig:EEG3}
\end{figure}

Figure \ref{fig:EEG3} shows the results of two typical ICs (with large energy) from the recovered dataset (Figure \ref{fig:EEG3} (b)(d)), and the results of corresponding ICs from the original dataset (Figure \ref{fig:EEG3} (a)(c)). Comparing Figure \ref{fig:EEG3} (a) with (b) and Figure \ref{fig:EEG3} (c) with (d) reveals that there is little difference in terms of scalp maps, ERP images, and averaged ERPs. This implies that BSBL-BO can recover EEG signals with satisfactory quality,  ensuring subsequent signal analysis with high fidelity.

\subsection{Experiment 2: Compressed Sensing with WT}

The second experiment used the dataset in \cite{wang2009predicting}. It consists of multiple channel signals, each channel signal containing 250 epochs for each of two events (`left direction' and `right direction'). Each epoch consists of 256 sampling points. The goal in \cite{wang2009predicting} is to differentiate the averaged ERP for the `left direction' with the averaged ERP for the `right direction'. For simplicity, we randomly chose a channel signal from the left parietal area. BSBL-BO and the previous $\ell_1$ algorithm were compared. The sensing matrix $\mathbf{\Phi}$ had the size of $128\times 256$, and the dictionary matrix $\mathbf{D}$ had the size of $256 \times 256$.

\begin{figure}[t]
	\begin{center}
	\subfigure[$\ell_1$]{
	\epsfig{figure=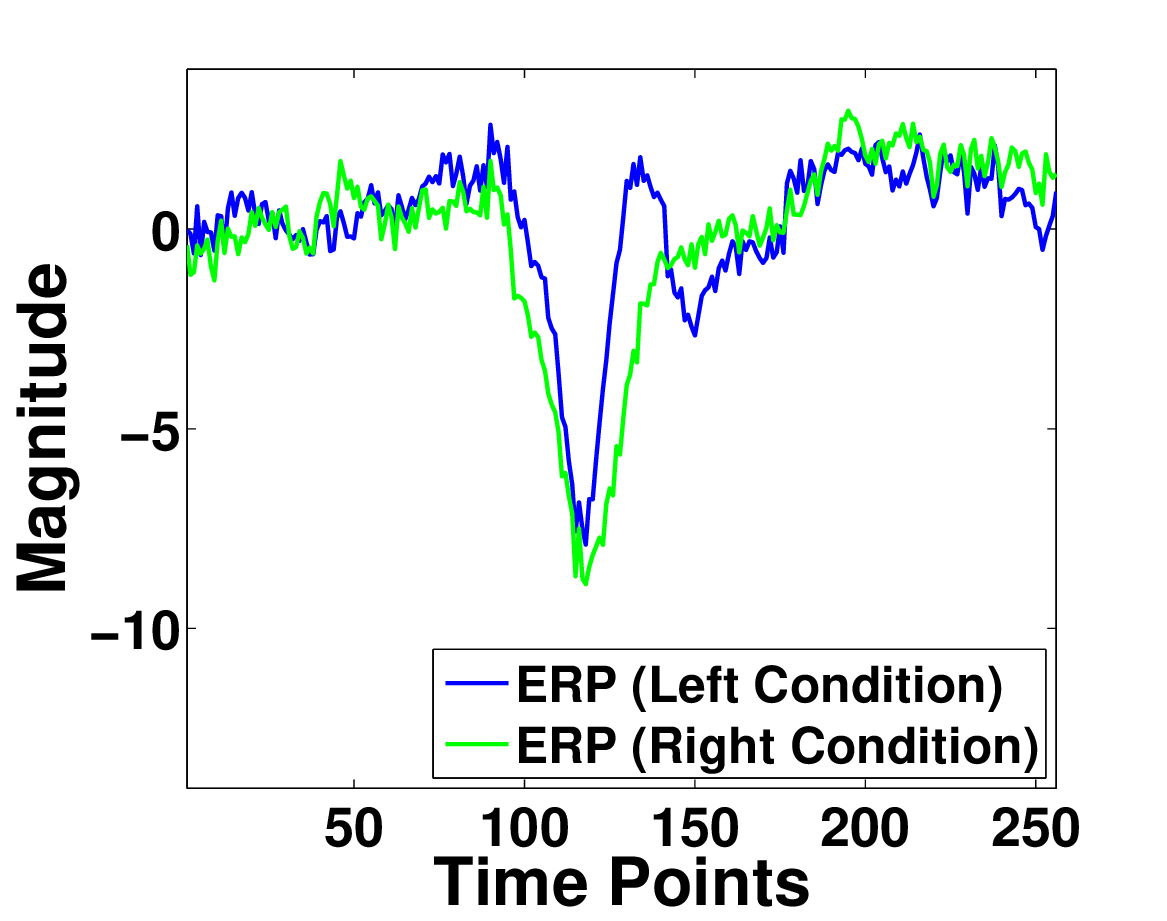,width=2.75cm,height=2.9cm}}
	\subfigure[BSBL-BO]{
	\epsfig{figure=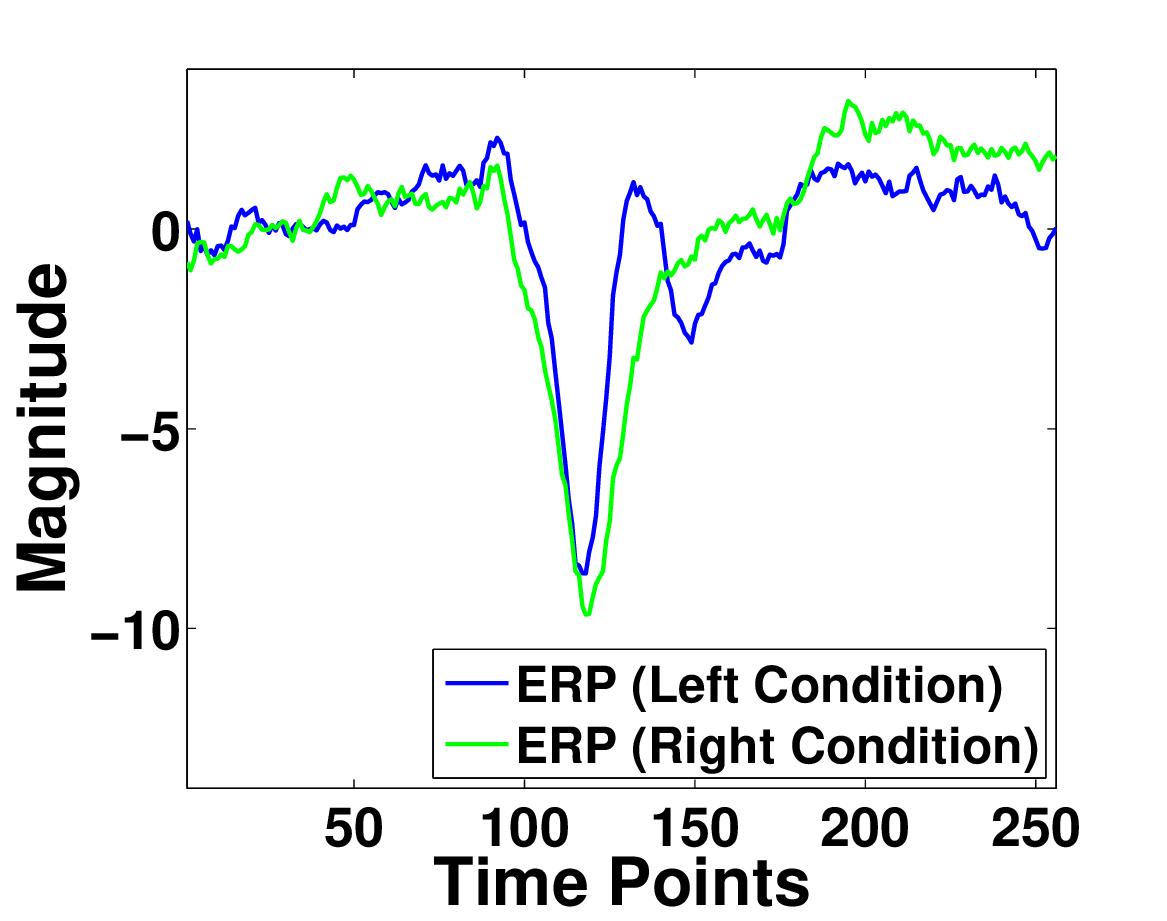,width=2.75cm,height=2.9cm}}
	\subfigure[Original Dataset]{
	\epsfig{figure=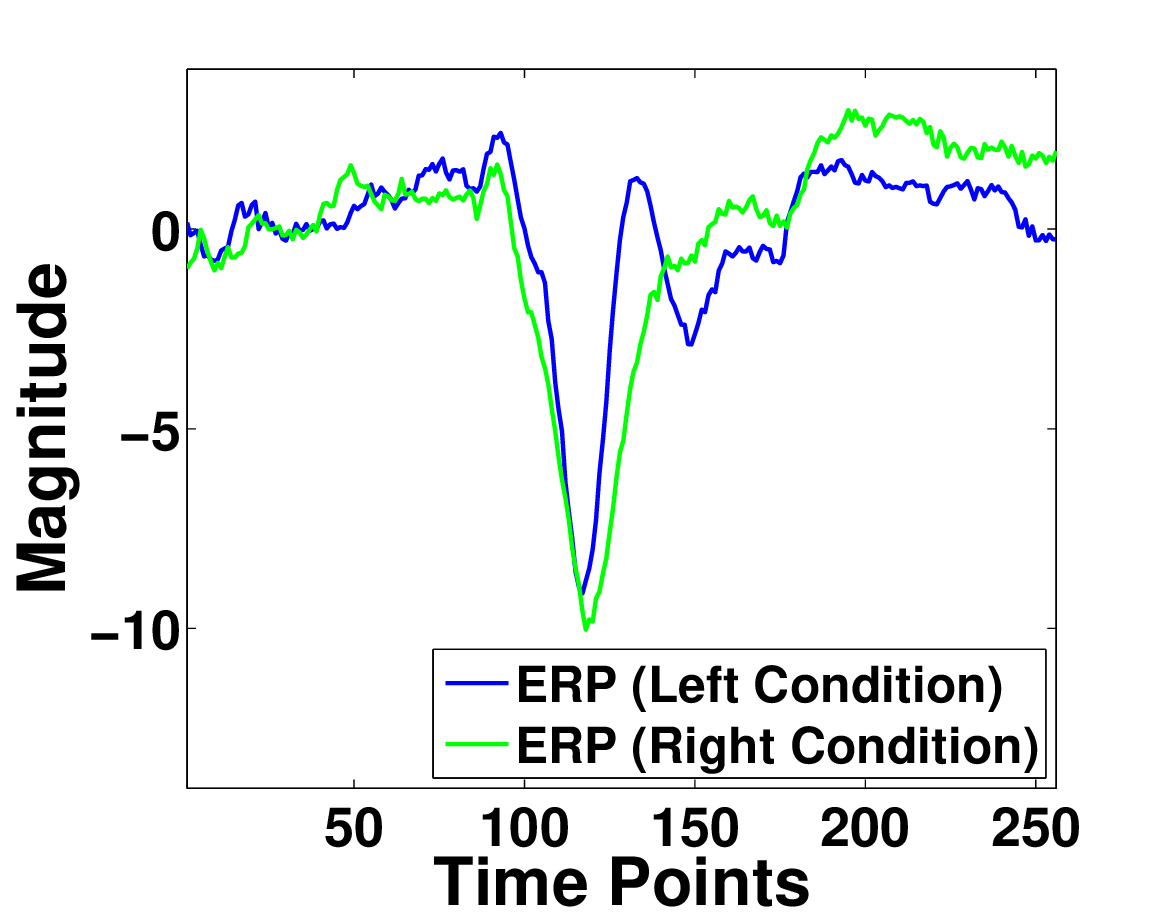,width=2.75cm,height=2.9cm}}
	\end{center}
	\caption{The ERPs corresponding to two event conditions (`left' and `right') averaged (a) from the recovered epochs by the $\ell_1$ algorithm, (b) from the recovered epochs by BSBL-BO, and (c) from the original dataset.}
	\label{fig:discussEEG2}
\end{figure}

For each event, we calculated the ERP by averaging the associated 250 recovered epochs. Figure \ref{fig:discussEEG2} (a) shows the ERP for the `left direction' and the ERP for the `right direction' averaged from the  dataset recovered by the $\ell_1$ algorithm.  Figure \ref{fig:discussEEG2} (b) shows the two ERPs averaged from the recovered dataset by BSBL-BO. Figure \ref{fig:discussEEG2} (c) shows the averaged ERPs from the original dataset (called genuine ERPs). Clearly, the resulting ERPs by the  $\ell_1$ algorithm were noisy. Although they maintained the main peaks of both genuine ERPs, they did not maintain other details of the genuine ERPs. Particularly, the difference between the two resulting ERPs from the $160^{\mathrm{th}}$ to the $250^{\mathrm{th}}$ time points was not clear. Besides, we found there were many brief oscillatory bursts in the recovered epochs by the $\ell_1$ algorithm (due to space limit we omit the results here). In contrast, the ERPs averaged from the recovered epochs by BSBL-BO maintained all the details of the genuine ERPs with high fidelity.

The SSIM and the NMSE of the resulting ERPs by the $\ell_1$ algorithm were 0.92 and 0.044, respectively. In contrast, the SSIM and the NMSE of the resulting ERPs by BSBL-BO were 0.97 and 0.008, respectively. In the experiment BSBL-BO took 0.06 second per epoch on average on the previous computer.

%============================================================================================
\section{Discussions}
\label{sec:discussions}
%============================================================================================
CS often resorts to a dictionary matrix to recover a non-sparse signal. However, the success of this approach heavily relies on the sparsity of its representation coefficients under the dictionary matrix. Therefore, finding a  dictionary matrix under which a signal can be sparsely represented is very important. However, finding such a dictionary matrix for many physiological signals is challenging. We found that using various popular dictionary matrices, the representation coefficients of EEG signals are still not sparse. Therefore, current CS algorithms have poor performance, and their recovery quality is not suitable for many clinical applications and cognitive neuroscience studies. Instead of seeking optimal dictionary matrices, this study proposed a method using general dictionary matrices yet achieving sufficient recovery quality for typical cognitive neuroscience studies. The empirical results suggest that when using the BSBL framework for EEG compression/recovery, the seeking of optimal dictionary matrices is not very crucial.

Notably, if energy consumption is not an issue, using wavelet compression for EEG can result in better recovery quality than any CS algorithms including BSBL-BO. For example, in the second experiment when transmitting 128 largest wavelet coefficients (using the same wavelet transform), we obtained the ERPs with higher quality: the SSIM and the NMSE were 0.99 and 0.0003, respectively. Clearly, if energy consumption is not a problem, wavelet compression may be a better choice. But sometimes higher recovery quality is not necessarily required for practical applications. Considering the ERP analysis in the second experiment, the recovery quality by BSBL-BO satisfied the requirement of the ERP analysis, and thus higher recovery quality is not needed or attractive, especially at the cost of more energy consumption. Of course, the choice between CS and wavelet compression (or other compression techniques) for the telemonitoring of EEG  probably differs case by case and thus needs further study.

%============================================================================================
\section{Conclusions}
\label{sec:conclusions}
%============================================================================================

Compressing EEG for telemonitoring is extremely difficult for current CS algorithms, because EEG is not sparse in the time domain nor sparse in transformed domains. To alleviate the problem, this study proposed to use the framework of block sparse Bayesian learning, which has superior performance to other existing CS algorithms in recovering non-sparse signals. Experimental results showed that it recovered EEG signals with good quality, ensuring subsequent signal analysis. Thus, it is very promising for wireless telemonitoring based cognitive neuroscience studies and engineering applications.

\bibliographystyle{IEEEtran}

\bibliography{zhilin,ecg,eeg,sparse}

\end{document}